\documentclass[prl,twocolumn,superscriptaddress]{revtex4}
\usepackage{epsfig}
\usepackage{times}
\usepackage{amsmath}
\usepackage{amsfonts}
\usepackage{amssymb}
\usepackage[usenames]{color}
\usepackage{graphicx}
\newcommand{\NTT}{NTT Basic Research Laboratories, NTT Corporation, 3-1 Morinosato-Wakamiya, Atsugi, Kanagawa, 243-0198, Japan.}

\newcommand{\keio}{Department of Applied Physics and Department of Physico-Informatics, Faculty of Science and Technology,
Keio University, Hiyoshi, Kohoku-ku, Yokohama 223-8522, Japan}

\newcommand{\beq}{\begin{equation}}
\newcommand{\eeq}{\end{equation}}
\newcommand{\beqa}{\begin{eqnarray}}
\newcommand{\eeqa}{\end{eqnarray}}

\renewcommand{\vec}[1]{\ensuremath{\mathbf{#1}}}

\begin{document}
\title{Single spin detection
with an ensemble of probe spins
}
\author{ Syuhei Uesugi}
   \affiliation{
\keio
   }
\author{Yuichiro Matsuzaki}
   \affiliation{
\NTT
   }
 \author{Suguru Endo
 }
 \affiliation{
\keio
   }
    \author{Shiro Saito}
   \affiliation{
\NTT
   }
 \author{Junko Ishi-Hayase
 }
 \affiliation{
\keio
   }

\begin{abstract}
Single spin detection is a key objective in the field of
 metrology.
There have been many experimental and theoretical investigations for the
 spin detection based on the use of probe spins.
 A probe 
 spin shows {{the}} precession due to {{dipole-dipole}} interaction from  {{a}}
 target spin, and measurement results of the probe spin
  {{allow}} us  {{to}}
 estimate the state of the target spin.
 Here, we investigate performance of single-spin detection when
 {{using}} an ensemble of probe spins.
  {{Even though}} the ensemble of probe  {{spins}} inevitably induces projection noise
 that could hinder the signal from the target spin, optimization of the configuration of the spin ensemble improves
 the sensitivity  {{such that}} enhancement of the signal can be much larger
 than the projection noise.
 The probe-spin ensemble is especially useful  {{at}} a large distance
 from the target spin  {{, where}} it is difficult for a single spin to read out
 the target spin within a reasonable repetition time.
 Our results pave the way for a new strategy
 to realize efficient  {{single-spin detections}}. 
\end{abstract}

\maketitle

 {{An}} important objective in quantum metrology is to realize
the efficient detection of a single spin. This technique has numerous potential
applications because we can in principle extract useful information about
materials by imaging nuclear magnetism on the nanometer scale.
However, such single-spin detection  {{requires}} both sensitivity and spatial
resolution.
There have been  {{several}} experimental and theoretical  {{studies}} to improve
both the sensitivity and spatial  {{resolution of}}
magnetic-field sensors such as SQUID, a superconducting flux qubit, Hall
sensors, and force sensors \cite{ramsden2006hall,vasyukov2013scanning,bal2012ultrasensitive,toida2016electron,bienfait2016reaching}.
 {{Even though}} there are some experimental demonstrations of single-spin
detection \cite{rugar2004single},
single-spin detection is not  {{yet}} a mature technology
.
 {{This is a particularly true because many repetitions}} of the measurements are necessary to increase the
signal {{-}}to {{-}}noise ratio for the spin detection  {{and}} much more efficient schemes are
required to realize rapid spin detection.

  {{The}} use of a probe spin
is one attractive approach for the single-spin detection \cite{degen2008microscopy,maze2008nanoscale,
taylor2008high, balasubramanian2008nanoscale, schaffry2011proposed,muller2014nuclear,staudacher2013nuclear,mamin2013nanoscale,ohashi2013negatively,rugar2015proton}.
 The probe spin can be coupled with
 the target spin via dipole-dipole interaction, and the probe spin  {{experiences}} a precession  {{due to}} the magnetic field
  {{induced by}} the target spin.
 From  {{an optical or electrical}} readout  {{of}} the state of the probe spin , we can estimate  {{the magnitude of the}} magnetic field applied
  {{to}} the probe spin, which provides us with information  {{on}} the target spin.

 {{Conversely}}, there have been several theoretical and experimental
studies of  {{ensembles}} of spins  {{for use as}} sensitive magnetic-field
sensors \cite{vengalattore2007high,kominis2003subfemtotesla,acosta2009diamonds,maertz2010vector,le2013optical,wolf2015subpicotesla}.
 If we use an ensemble of
spins to measure  {{the}} applied magnetic fields,
we can enhance
the signal from the target magnetic field. This  {{has}} a clear advantage over  {{a}} single-
spin field sensor if we aim to detect global magnetic fields.
However,  {{an}} ensemble of  {{spins has}}
projection noise  {{resulting}} from the intrinsic properties of quantum
mechanics where the readout of the quantum states becomes a stochastic
process.
This problem could be significant if we aim to use the
probe-spin ensemble to detect a single target spin.
The dipole-dipole interaction between spins has  {{the}} form  $H_{\rm{dd}}\propto
1/r^3$ where $r$ denotes the distance between the spins {{; therefore,}}
the interaction becomes significantly weaker as we increase the distance
from the target spin. This means that, if we use  {{an}} ensemble of probe spins to detect the
target spin, probe spins far from the target spins could 
induce projection noise without  {{contributing to}} the enhancement
of the signal.
 {{Therefore}},
a careful assessment is required to
determine the conditions when  {{a}} probe-spin ensemble shows better performance
than  {{a}} single probe spin. 

In this paper, we  {{investigate}} performance of the single-spin
detection with  {{a}} probe-spin ensemble.
Interestingly, we found that, by choosing  {{a}} suitable distribution of
probe spins, the use of  {{a}} probe-spin ensemble is much more
efficient than that of  {{a}} single probe spin.
As a concrete example, we consider nitrogen vacancy centers. By
performing numerical simulations with realistic parameters, we 
found that the sensitivity of the probe-spin ensemble becomes more than
10 times better than that of the single probe spin.

The remainder of this paper is organized as follows.
In Sec. II we review magnetic-field sensing with the
standard echo technique. In Sec. III we investigate the
performance of single-spin detection using a probe spin.
In Sec. IV we
introduce a spin detection scheme  {{using}} an ensemble of probe spins. Finally, in Sec. V, we  {{offer}}
our  {{conclusions}}.

\section{Sensing global magnetic fields with a probe spin}
\begin{figure}[h] 
\begin{center}
\includegraphics[width=1.03\linewidth]{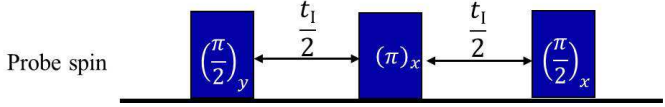} 
\caption{
  {{Pulse}} sequence to detect global AC magnetic fields  {{using the}} spin echo
 technique.
 The $\pi $ pulse in the middle improves the coherence time of the
 probe spin, because  {{it}} removes  {{the}} low frequency noise.
 }
 \label{echo}
\end{center}
\end{figure}

Let us review sensing global magnetic fields with
a single probe spin using the standard echo measurement.
The
Hamiltonian is described as
 \begin{eqnarray}
  H&=&\frac{\omega -\omega '}{2}\hat{\sigma }_z+\lambda 
   \hat{\sigma}_x\cos (\omega 't+\phi )\nonumber \\
 \end{eqnarray}
 where $\omega =g\mu _bB_{\rm{ex}}+g\mu _bB(t)$ denotes the resonant
 frequency of the probe spin,  $g$ denotes a g factor, $\mu _b$
 denotes a Bohr magneton, $B_{\rm{ex}}$ denotes a known external
 magnetic field, $B(t)$
denotes a target global magnetic field, $\omega '=g\mu _bB_{\rm{ex}}$  {{denotes the}}  frequency of
the microwave fields, $\lambda $ denotes a Rabi frequency, and $\phi $
denotes  {{the}} phase of the microwave fields.
After  {{applying}} a rotating wave
approximation, we obtain
 \begin{eqnarray}
  H&\simeq &\frac{g\mu _bB(t)}{2}\hat{\sigma }_z+\lambda _x \hat{\sigma
   }_x+\lambda _y \hat{\sigma
   }_y\nonumber \\
 \end{eqnarray}
where $\lambda _x$ ($\lambda _y$) denotes a Rabi
frequency of the microwave along the $x$ ($y$) direction.
We turn off the microwave driving ($\lambda _x=\lambda _y=0$) except
when we need to rotate the probe spin.
 We define
$H_{\rm{FID}}=\frac{g\mu _bB(t)}{2}\hat{\sigma }_z$ for the
Hamiltonian without microwave driving.
 {{In particular}}, we consider alternating square fields ( {{which}} can be
considered  {{to be}} AC fields \cite{dolde2011electric}) described as
\begin{eqnarray}
B(t)=\left\{
\begin{array}{l}
 B \ \ (0\leq t <\frac{t_{\rm{I}}}{2})\\
 -B    (\frac{t_{\rm{I}}}{2}\leq t \leq t_{\rm{I}})
\end{array}
\right.
\end{eqnarray}
where $t_{\rm{I}}$ denotes an interaction between the probe spin and
 {{the}} target magnetic fields.
We describe a scheme to estimate the value of $B$ with the probe spin  {{at}} a
given time $T$ ( {{see}} Fig. \ref{echo}).
 {{First}}, we prepare a state of $ |\psi _0\rangle =|+\rangle =\frac{1}{\sqrt{2}}(|0\rangle +|1\rangle ) $ by
performing $\frac{\pi }{2}$ pulse along the $y$ direction.
 {{Second, we allow}} this state  {{to}} evolve with the Hamiltonian $H_{\rm{FID}}$ for
time $\frac{t_{\rm{I}}}{2}$.  {{Because}} the probe spins are affected by dephasing from the
environment, the non-diagonal terms of the density matrix 
decay.  {{Taking this}} decoherence  {{into consideration}}, the density matrix after the
evolution  {{at}} time $t$ is given as
$ \rho (t)=\frac{1}{2}|0\rangle \langle 0|+\frac{1}{2}e^{-i
g\mu _bB
  t-\gamma t}|1\rangle \langle 0|
  +\frac{1}{2}e^{i g\mu _bB
  t-\gamma t}|0\rangle \langle 1|+\frac{1}{2}|1\rangle \langle 1|$
where
$\gamma $ denotes  {{the}} dephasing rate.
 {{Third}}, after performing a $\pi $ pulse along the $x$ axis to flip the
probe spin at time $t=\frac{t_{\rm{I}}}{2}$, we  {{allow}} this state  {{to}} evolve with the Hamiltonian $H_{\rm{FID}}$ for
time $\frac{t_{\rm{I}}}{2}$.  {{Note}} that this $\pi $
pulse at $t=\frac{t_{\rm{I}}}{2}$ suppresses  {{the}} low-frequency fluctuations of the resonant frequency of the
probe spin, which improves the coherence time \cite{de2010universal}.
 {{Fourth}}, we perform a projective measurement on this state about an
observable $\hat{\sigma }_y$, which can be realized by a $\hat{\sigma
}_z$ measurement after  {{a}} $\frac{\pi}{2}$ pulse along the $x$ direction.
The expectation value is calculated as
$ \langle \hat{\sigma}_y\rangle =e^{-\gamma t_{\rm{I}}}\sin
g\mu _bB
  t_{\rm{I}}\simeq e^{-\gamma t}  g\mu _bB t_{\rm{I}}$
where we use $\omega t_{\rm{I}}\ll 1$.
Finally, we repeat the above three steps within a given time $T$. We
assume that the necessary time for the single qubit rotation and  {{the}}
measurements  {{is}} much shorter than the coherence time of the probe
spins. In this case, the number of trials  {{in}} a given time $T$ is
approximated as $N\simeq T/t_{\rm{I}}$.
We can calculate the uncertainty of the estimation of the magnetic
fields as follows {{:}}
\begin{eqnarray}
 \delta B&=&\frac{\sqrt{\langle \delta \hat{\sigma }_y\delta
  \hat{\sigma}_y\rangle}}{|\frac{d\langle
  \hat{\sigma}_y\rangle}{dB}|}\frac{1}{\sqrt{N}}\nonumber \\
\end{eqnarray}
where $\delta \hat{\sigma}_y\equiv \hat{\sigma}_y -\langle \hat{\sigma}_y\rangle $.
If the magnetic fields are small, we can simplify the uncertainty as
$\delta B\simeq \frac{1}{e^{- t_{\rm{I}}/T_2}g\mu _b
t_{\rm{I}}}\frac{1}{\sqrt{\frac{T}{t_{\rm{I}}}}}$ where $T_2=\frac{1}{\gamma }$ denotes  {{the}}
coherence time.
We can minimize this uncertainty by choosing $t=\frac{T_2 }{2}$, and
 {{thus}} obtain
 $ \delta B_{\rm{min}}\simeq \frac{1}{e^{-\frac{1}{2}}g\mu _b
   }\frac{1}{\sqrt{\frac{T_2T}{2 } }}$.

\section{Single-spin detection using a single probe spin.}
\begin{figure}[h] 
\begin{center}
\includegraphics[width=0.9\linewidth]{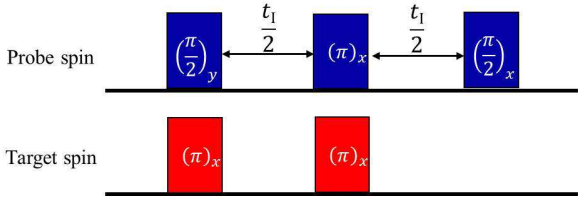} 
\caption{
  {{Pulse}} sequence to detect the state of the target spin with  {{a}}
 probe spin. The pulse sequence on the probe spin is the same as
 that used to detect  {{the}} AC magnetic fields, as described in 
 Fig. \ref{echo}.
  {{Note}} that, in order to generate effective AC
 magnetic fields from the target spin, we perform two $\pi $ pulses on the
 target spin \cite{mamin2013nanoscale}.
 }
 \label{double}
\end{center}
\end{figure}

 We  {{now}} consider  {{detecting}} a target spin using a single probe spin.
 When the target spin is located at the origin of  {{the coordinate system}}, the Hamiltonian between the target spin and  {{the}} probe spin is described
 as follows {{:}}
 \begin{eqnarray}
  H&=&H_{\rm{e}}+H_{\rm{I}}+H_{\rm{t}} \nonumber \\
  H_{\rm{e}}&=&\frac{g_{\rm{e}}\mu _bB_{\rm{ex}}}{2}\hat{\sigma
   }^{(\rm{e})}_z +\lambda ^{(\rm{e})} \hat{\sigma
   }^{(\rm{e})}_x \cos (\omega 't +\phi ')
   \nonumber \\
  H_{\rm{I}}&=&G\frac{1}{|\vec{r}|^3}\Big{(}3\frac{1}{|\vec{r}^2|}(\vec{\sigma
   ^{(e)}}\cdot \vec{r})(\vec{\sigma
   ^{(t)}}\cdot \vec{r})-(\vec{\sigma
   ^{(e)}}\cdot \vec{\sigma
   ^{(t)}})
   \Big{)}
   \nonumber \\
    H_{\rm{t}}&=&\frac{g_{\rm{t}}\mu _bB_{\rm{ex}}}{2}\hat{\sigma
     }^{(\rm{t})}_z+\lambda ^{(\rm{t})} \hat{\sigma
   }^{(\rm{t})}_x \cos (\omega ''t +\phi '')
     \nonumber \\
 \end{eqnarray}
 where $B_{\rm{ex}}$ denotes  {{the}} external magnetic fields,
 $\lambda^{(\rm{e})}$($\lambda^{(\rm{t})}$) denotes a Rabi frequency for
 the probe (target) spin, $\omega '=g_{\rm{e}}\mu _bB_{\rm{ex}}$
 ($\omega ''=g_{\rm{t}}\mu _bB_{\rm{ex}}$) denotes  {{the}} frequency of the
 microwave fields on the probe (target) spin, $\phi '$ ($\phi ''$)
 denotes the phase of the microwave fields, and $\vec{r}=(x,y,z)$
 denotes  {{the}} position of the probe spin. 
  {{In addition}}, we have $\vec{\sigma
   ^{(e)}}\cdot \vec{r}=x\hat{\sigma }^{(\rm{e})}_x+y\hat{\sigma
 }^{(\rm{e})}_y+z\hat{\sigma }^{(\rm{e})}_z$ and $\vec{\sigma
   ^{(t)}}\cdot \vec{r}=x\hat{\sigma }^{(\rm{t})}_x+y\hat{\sigma }^{(\rm{t})}_y+z\hat{\sigma }^{(\rm{t})}_z$.
For $g_{\rm{e}}\mu _bB_{\rm{ex}}\gg g_{\rm{t}}\mu _bB_{\rm{ex}}$, we
use a rotating wave approximation, and simplify the Hamiltonian
in a rotating frame
as
 \begin{eqnarray}
  H_{\rm{e}}&\simeq&\lambda _x^{(\rm{e})} \hat{\sigma
   }^{(\rm{e})}_x +\lambda ^{(\rm{e})}_y \hat{\sigma
   }^{(\rm{e})}_y
   \nonumber \\
  H_{\rm{I}} &\simeq &
   \frac{G}{(x^2+y^2+z^2)^{\frac{3}{2}}}(\frac{3z^2}{x^2+y^2+z^2}-1)\hat{\sigma
   }^{(\rm{e})}_z\hat{\sigma
   }^{(\rm{t})}_z
   \nonumber \\
  H_{\rm{t}} &\simeq&\lambda _x^{(\rm{t})} \hat{\sigma
   }^{(\rm{e})}_x +\lambda ^{(\rm{t})}_y \hat{\sigma
   }^{(\rm{e})}_y
 \end{eqnarray}
 where $\lambda _x^{(\rm{e})}$ ($\lambda _x^{(\rm{t})}$) denotes a Rabi
  frequency along  {{the}} $x$ component on the probe (target) spin and
  $\lambda _y^{(\rm{e})}$ ($\lambda _y^{(\rm{t})}$) denotes a Rabi
  frequency along  {{the}} $y$ component on the probe (target) spin.
  We turn off the microwave driving ($\lambda _x^{(\rm{e})}=\lambda _x^{(\rm{t})}=\lambda _y^{(\rm{t})}=\lambda _y^{(\rm{t})} =0$) except
when we need to rotate the spins.
   {{Because}} the target is the spin $\frac{1}{2}$, the observable $\hat{\sigma}^{(\rm{t})}_z$ provides us with $+1$ or $-1$ depending on the target
    state after the measurement.
We can consider an effective Hamiltonian {{,}}
 \begin{eqnarray}
 H^{(\rm{eff})}_{\rm{I}}\simeq
  \frac{G}{(x^2+y^2+z^2)^{\frac{3}{2}}}(\frac{3z^2}{x^2+y^2+z^2}-1)s\hat{\sigma
  }^{(\rm{e})}_z
\end{eqnarray}
where $\hat{\sigma
  }^{(\rm{t})}_z$ is replaced  {{by}} a classical parameter $s $.
  In this case, the dipole-dipole interaction from the target spin can
  be treated as  {{the}} magnetic fields on the probe spin where the effective
  Zeeman splitting is defined as
  \begin{eqnarray}
B^{\rm{(eff)}}=\frac{2G}{   g_{\rm{e}}\mu _b(x^2+y^2+z^2)^{\frac{3}{2}}}(\frac{3z^2}{x^2+y^2+z^2}-1)s
  \end{eqnarray}
  Similar to the global magnetic-field sensing described above, we
  can estimate the parameter $s$  {{using}} the standard echo
  measurement where we flip the target spin in the middle to induce effective
  AC magnetic fields, as described in Fig. \ref{double}.
  If we have $s\simeq 1$ ($s\simeq -1$) as the estimated value, we  {{can}} conclude that the state of
  the target spin is up (down). 
Similar to the detection of the global magnetic fields, we can calculate
the uncertainty of the estimation of $s$ as follows {{:}}
\begin{eqnarray}
 \delta s ^{\rm{(single)}}
  \simeq
  \frac{1}{e^{-\frac{t_{\rm{I}}}{T^{(\rm{single})}_2}}\frac{2G}{(x^2+y^2+z^2)^{\frac{3}{2}}}|\frac{3z^2}{x^2+y^2+z^2}-1|
  t_{\rm{I}}\sqrt{\frac{T}{t_{\rm{I}}}}} 
\end{eqnarray}
where
 $T^{(\rm{single})}_2$ denotes  {{the}} coherence time of the single probe spin.
We can minimize this uncertainty
 by choosing $t_{\rm{I}}=\frac{1}{2 }T^{(\rm{single})}_2$, and
obtain
\begin{eqnarray}
 \delta s^{(\rm{single})}_{\rm{min}}\simeq \frac{1}{\frac{2e^{-\frac{1}{2}}G}{(x^2+y^2+z^2)^{\frac{3}{2}}}|\frac{3z^2}{x^2+y^2+z^2}-1|
  \sqrt{\frac{TT^{\rm{(single)}}_2}{2 }}}.\label{single}
\end{eqnarray}
 {{Note}} that we  {{need to}} decrease this uncertainty  {{to}} much smaller
than $1$ to determine the state of the target spin.

\section{Single-spin detection using an ensemble of probe spins}
 {{Here,we}} describe our scheme to detect the target spin with an ensemble of
probe spins.
In a rotating frame, the effective interaction Hamiltonian between the
probe spins and the target spin
is given as
\begin{eqnarray}
 H^{\rm{(eff)}}_{\rm{I}} &\simeq &\sum_{j=1}^{L}\frac{g_{\rm{e}}\mu _bB_j}{2}\hat{\sigma
   }^{(\rm{e})}_{z,j}
\end{eqnarray}
 \begin{eqnarray}
   B_j&= &
   2\frac{G}{g_{\rm{e}}\mu _b(x_j^2+y_j^2+z_j^2)^{\frac{3}{2}}}(\frac{3z_j^2}{x_j^2+y_j^2+z_j^2}-1)
   s
\nonumber 
 \end{eqnarray}
where $B_j$ denotes  {{the}} effective magnetic fields on the $j$ th probe
spins from the target spin, $\vec{r}_j=(x_j,y_j,z_j)$ denotes the position of the $j$ th
probe spin, $L$ denotes the number of probe spins,and $s$ denotes the state
of the target spin.
We use the same pulse sequence described in Fig. \ref{double}, and
assume that we can uniformly implement both  {{the}}  $\frac{\pi }{2}$
pulse and  {{the}}  $\pi $ pulse on the all the probe spins.
The uncertainty of the estimation for the probe spin ensemble
can be calculated as
\begin{eqnarray}
 \delta s^{(\rm{ens})}&=&\frac{\sqrt{\langle \delta \hat{M }_y\delta
  \hat{M}_y\rangle}}{|\frac{d\langle
  \hat{M}_y\rangle}{ds}|}\frac{1}{\sqrt{N}}\nonumber 
\end{eqnarray}
where $\hat{M}_y=\sum_{j=1}^{L}\hat{\sigma }_y^{(j)}$.
We obtain
 $\langle \hat{M}_y\rangle = \sum_{j=1}^{L}e^{-\frac{t_{\rm{I}}}{T^{\rm{(ens)}}_2}}\sin
   (g_{\rm{e}}\mu _b B_jt_{\rm{I}})
  \simeq
  \sum_{j=1}^{L}e^{-\frac{t_{\rm{I}}}{T^{\rm{(ens)}}_2}}g_{\rm{e}}\mu _b
  B_jt_{\rm{I}}$
  where $T_2^{\rm{(ens)}}$ denotes  {{the}} coherence time of the probe-spin
  ensemble.
 {{Taking}} a continuous limit, we obtain
\begin{eqnarray}
 \langle \hat{M}_y\rangle 
 \simeq 2G\rho t_{\rm{I}}e^{-\frac{t_{\rm{I}}}{T^{(\rm{ens})}_2}}s \int \int \int
 dxdydz
 \frac{(\frac{3z^2}{x^2+y^2+z^2}-1)}{(x^2+y^2+z^2)^{\frac{3}{2}}}\nonumber
\end{eqnarray}
 where
 we perform the integral over the region of the probe-spin ensemble
 by considering the spin density $\rho $.
 {{In addition}}, we obtain $\langle \delta \hat{M}_y\delta \hat{M}_y\rangle
=\sum_{j=1}^{L}\langle \delta \hat{\sigma }_y\delta \hat{\sigma
}_y\rangle\simeq L$ for  {{the}} small effective magnetic fields.
 {{Therefore,}} we obtain
\begin{eqnarray}
 \delta s^{(\rm{ens})}_{\rm{min}}&\simeq &\frac{\sqrt{L}}{|
  2e^{-\frac{1}{2}}G\rho  \int \int \int
 dxdydz
 \frac{(\frac{3z^2}{x^2+y^2+z^2}-1)}{(x^2+y^2+z^2)^{\frac{3}{2}}}
  |}\frac{1}{\sqrt{\frac{T^{(\rm{ens})}_2T}{2}}}.\nonumber 
\end{eqnarray}
where we choose $t_I=\frac{T^{\rm{(ens)}}_2}{2}$ to minimize the uncertainty.
 {{Because}} this form contains an integral over the  {{location}} where the probe
spins exist, we need to specify the shape and volume of the region of
the probe spins, as we will describe in the following subsections.

\subsection{Columnar form for the  {{distribution}} of probe spins}
  \begin{figure}[h!] 
\begin{center}
\includegraphics[width=0.9\linewidth]{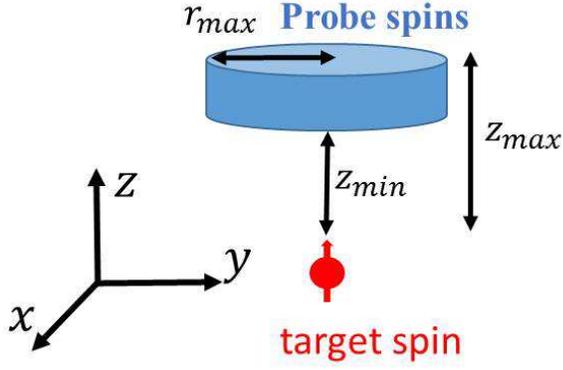} 
\caption{
 Detection of a target spin with an ensemble of probe spins. Here, we
 assume that the probe spins  {{are}} homogeneously distributed inside a columnar
 form that is placed at a distance from the target spin.
 }
 \label{column}
\end{center}
\end{figure}
First, we consider a columnar form for the distribution of the
probe spins, as shown in Fig. \ref{column}.
 {{Note}} that existing technology allows us to fabricate such a structure by combining 
electron-beam lithography and reactive ion etching, and we can
use this structure as the tip for a scanning microscope
\cite{maletinsky2012robust}.

  \begin{figure}[h] 
\begin{center}
\includegraphics[width=0.9\linewidth]{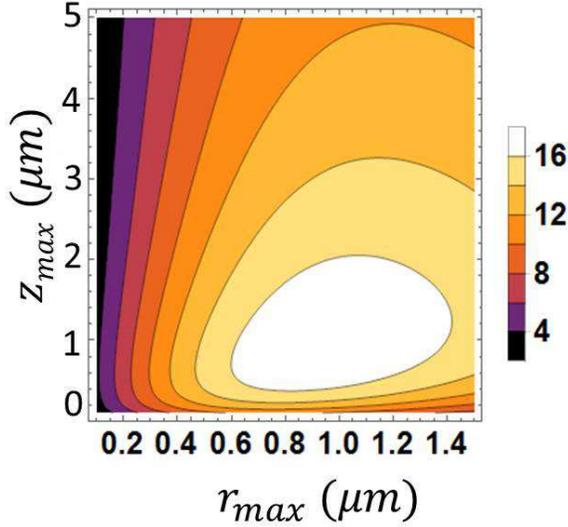} 
\caption{ {{Plot of the}} ratio $\delta s^{(\rm{single})}_{\rm{min}}/\delta
  s^{(\rm{ens})}_{\rm{min}}$ with a columnar configuration, as described in
  Fig. \ref{column}.
 We chose the parameters  $T_2^{(\rm{single})}=2$ ms
for the single probe spin, and $T_2^{(\rm{ens})}=84\,\mu $s and $\rho =6.7\times
10^{16}/{\rm{cm}^3}$ for the ensemble of the probe spins.  {{In addition}}, we  {{fixed}}
 $z_{\rm{min}}=1\,\mu $m.
 The ratio shows a maximum value of $\delta s^{(\rm{single})}_{\rm{min}}/\delta
  s^{(\rm{ens})}_{\rm{min}}\simeq 17.5$
 for $r_{\rm{max}}=0.93\,\mu $m and $z_{\rm{max}}=1.87 \mu m$.
 }
 \label{colthreed}
\end{center}
\end{figure}

We can calculate
\begin{eqnarray}
 &&  \int
 _0^{r_{\rm{max}}}dr\int _0^{2\pi } d\theta \int_{z_{\rm{min}}}^{z_{\rm{max}}}
 dz
 \frac{r(\frac{3z^2}{r^2+z^2}-1)}{(r^2+z^2)^{\frac{3}{2}}}\nonumber \\
 &=&
  2\pi 
  \Big{(}\frac{z_{\rm{max}}}{\sqrt{r^2_{\rm{max}}+z^2_{\rm{max}} }}
  -\frac{z_{\rm{min}}}{\sqrt{r^2_{\rm{max}}+z^2_{\rm{min}} }}
  \Big{)}
  \nonumber 
\end{eqnarray}
where $\rho $ denotes the density of the probe spin.
Therefore, the uncertainty of the estimation is calculated as
\begin{eqnarray}
 \delta s^{(\rm{ens})}_{\rm{min}}\simeq \frac{\sqrt{\rho (z_{\rm{max}}-z_{\rm{min}})\pi r^2_{\rm{max}}}}{4\pi \rho Ge^{-\frac{1}{2}}
  \Big{(}\frac{z_{\rm{max}}}{\sqrt{r^2_{\rm{max}}+z^2_{\rm{max}} }}
  -\frac{z_{\rm{min}}}{\sqrt{r^2_{\rm{max}}+z^2_{\rm{min}} }}
  \Big{)}}\frac{1}{\sqrt{\frac{T_2T}{2}}}\nonumber
\end{eqnarray}
For comparison, we consider  {{the}} uncertainty when we use a single
probe spin for the spin detection by substituting
$x=y=0$ and $z=z_{\rm{min}}$ into
Eq. \ref{single}.
\begin{eqnarray}
 \delta s ^{\rm{(single)}}
  \simeq
  \frac{1}{e^{-\frac{1}{2}}\frac{4G}{z^3_{\rm{min}}}}\frac{1}{\sqrt{\frac{T_2T}{2}}}
\end{eqnarray}
We can define the ratio of the uncertainty of the estimation as
\begin{eqnarray}
 \frac{\delta s^{(\rm{single})}_{\rm{min}}}{\delta
  s^{(\rm{ens})}_{\rm{min}}}=
  \frac{\pi \sqrt{\rho} z^3_{\rm{min}}
  \Big{(}\frac{z_{\rm{max}}}{\sqrt{r^2_{\rm{max}}+z^2_{\rm{max}} }}
  -\frac{z_{\rm{min}}}{\sqrt{r^2_{\rm{max}}+z^2_{\rm{min}} }}
  \Big{)}}{\sqrt{ (z_{\rm{max}}-z_{\rm{min}})\pi r^2_{\rm{max}}}}\sqrt{\frac{T_2^{(\rm{ens})}}{T_2^{\rm{(single)}}}}\nonumber
\end{eqnarray}
To calculate this ratio,
 we performed numerical
simulations.

  \begin{figure}[h] 
\begin{center}
\includegraphics[width=0.9\linewidth]{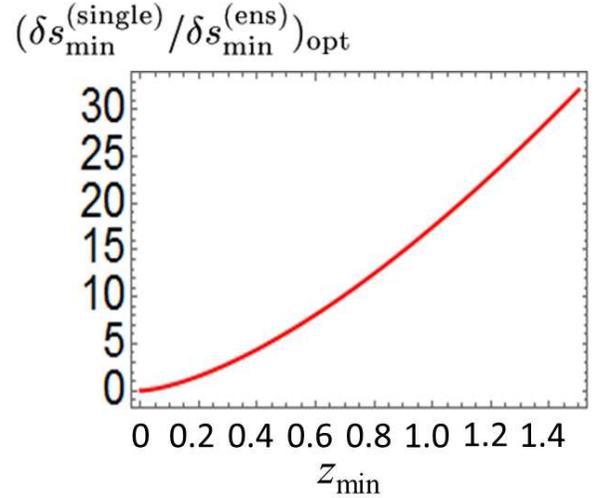} 
\caption{We plot an optimize ratio  $\delta s^{(\rm{single})}_{\rm{min}}/\delta
  s^{(\rm{ens})}_{\rm{min}}$ against $z_{\rm{min}}$ where we choose $r_{\rm{max}}$ and
 $z_{\rm{max}}$ to maximize this ratio by a continuous line. Except these two
 parameters, we used the same parameters as the Fig. \ref{colthreed}.
 The ensemble probe spins shows better performance than the single probe
 spin as long as $z_{\rm{min}}\geq 0.15 \mu $m.
 }
  \label{coltwod}
\end{center}
\end{figure}

For the simulations, we used typical parameters for  {{the}} nitrogen vacancy (NV)
centers in diamond.
The NV center is  {{a}} fascinating  {{candidate for realizing}} a sensitive  magnetic-field sensor \cite{maze2008nanoscale,
taylor2008high, balasubramanian2008nanoscale, schaffry2011proposed}.
We can use this system as an effective two-level system, and high
fidelity gate operations  {{using}} microwave pulses have  {{already}} been demonstrated
\cite{Go01a,gruber1997scanning,jelezko2002single,JGPGW01a}.
Moreover,  it is known that we can  {{read out}} the state of the NV centers
 {{via}} fluorescence
from the optical transitions after  {{irradiation with a}} green laser \cite{gruber1997scanning,jelezko2002single}.
 {{In particular,}} single NV centers have a long coherence time  {{, e.g.,}} a few  {{milliseconds}}
\cite{balasubramanian2009ultralong,mizuochi2009coherence}.
 {{It}} is possible to fabricate  high-density NV centers
 {{, which}} have been used for magnetic-field sensing
\cite{acosta2009diamonds,maertz2010vector,le2013optical,wolf2015subpicotesla}{{;however,}} the
coherence time of {{an}} ensemble of NV centers is typically much shorter
than that of {{a}} single NV center.
In our numerical simulations, we use values {{of}}  $T_2^{(\rm{single})}=2$ ms
for a single NV center {{and}}  $T_2^{(\rm{ens})}=84\,\mu $s and $\rho =6.7\times
10^{16}/{\rm{cm}^3}$ for an ensemble of NV centers
\cite{balasubramanian2009ultralong,grezes2015storage,wolf2015subpicotesla}.
{{Note}} that, {{even though}} we {{focused}} on NV centers {{in}} the
numerical simulations, we can{{,}} in principle{{,}} use other spin ensembles such as donors in high-
purity silicon or erbium impurities in yttrium orthosilicate, which can
be {{read out via a}} superconducting circuit
\cite{bushev2011ultralow,tyryshkin2012electron,tanaka2015proposed,toida2016electron,bienfait2016reaching}.

We plot the ratio $\delta s^{(\rm{single})}_{\rm{min}}/\delta
  s^{(\rm{ens})}_{\rm{min}}$ against $z_{\rm{max}}$ and $r_{\rm{max}}$
  in Fig. \ref{colthreed}, where we fix
 $z_{\rm{min}}=1\,\mu $m.
  There exists an optimal set of 
$z_{\rm{max}}$ and $r_{\rm{max}}$, and the maximized ratio is {{approximately}}
  $\delta s^{(\rm{single})}_{\rm{min}}/\delta
  s^{(\rm{ens})}_{\rm{min}}\simeq 17.5$. This means that the ensemble of
  {{spins}} actually shows  better performance for spin detection than a single probe spin
  with these realistic parameters.
  {{In addition}}, we {{plotted the}} optimized value of $\delta s^{(\rm{single})}_{\rm{min}}/\delta
  s^{(\rm{ens})}_{\rm{min}}$ against $z_{\rm{min}}$ in Fig. \ref{coltwod}
  ,where we chose
  $z_{\rm{max}}$ and $r_{\rm{max}}$ to maximize $\delta s^{(\rm{single})}_{\rm{min}}/\delta
  s^{(\rm{ens})}_{\rm{min}}$.{{The}} probe-spin ensemble
  {{has}} better sensitivity than the single probe spin as long as  $z_{\rm{min}}\geq 0.15\ \mu $m.

\subsection{Cylindrical form form for the distribution of probe spins}
  \begin{figure}[h!] 
\begin{center}
\includegraphics[width=0.7\linewidth]{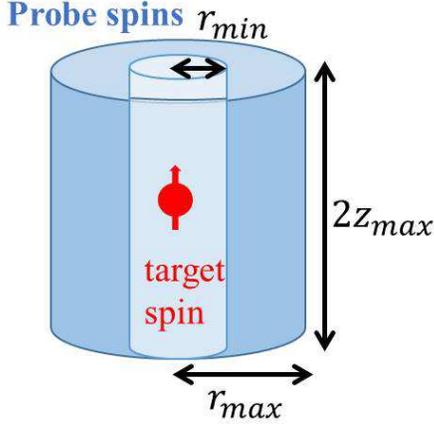} 
\caption{
 Detection of a target spin with an ensemble of probe spins using a
 cylindrical form. Here, after fabricating the probe-spin substrate into a columnar
 form with  radius of $r_{\rm{max}}$, we created a hole penetrating the structure with  radius  $r_{\rm{min}}$.
 We assume that the probe spins {{are}} homogeneously {{distributed}} in the
 substrate and {{that}} the target spin is located inside the hole.
 }
 \label{cylinder}
\end{center}
\end{figure}
{{Second}}, we consider a cylindrical form for the distribution of the
probe spins, as shown in Fig. \ref{cylinder}.
After we fabricate a probe-spin substrate (such as diamonds) into a
columnar form, we make a hole penetrating the structure{{.The}}  target
spin is located in the center of the hole.
Such a fabrication is possible if we use a focused ion beam
\cite{hadden2010strongly}.
Unlike the columnar form as described in the previous subsection, it is
difficult to
use this structure {{with}} a scanning microscope because the target spin is
assumed to be inside the cylindrical form.
However, as we will describe, this cylindrical form shows much better
performance than the columnar form for spin {{detections.Therefore}}, for a proof of principle experiment, this structure would be suitable.

  \begin{figure}[h!] 
\begin{center}
\includegraphics[width=0.9\linewidth]{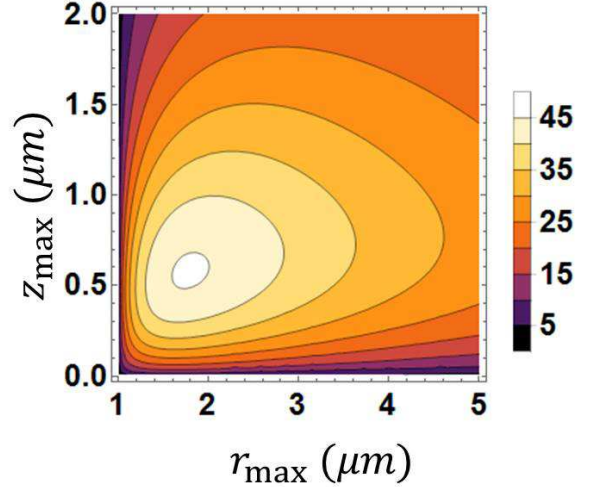} 
\caption{{{Plot of the}} $\delta s^{(\rm{single})}_{\rm{min}}/\delta
  s^{(\rm{ens})}_{\rm{min}}$ with a {{cylindrical}} configuration, as described in Fig. \ref{cylinder}. 
 We {{chose}} the same parameters as {{in}} Fig. \ref{cylthreed}.
 {{In addition}}, we {{fixed}}
 $r_{\rm{min}}=1\,\mu $m.
 The ratio shows a maximum value of $\delta s^{(\rm{single})}_{\rm{min}}/\delta
  s^{(\rm{ens})}_{\rm{min}}\simeq 45$
 for $r_{\rm{max}}=1.77\,\mu $m and $z_{\rm{max}}=0.58\,\mu m$.
 }
 \label{cylthreed}
\end{center}
\end{figure}

  \begin{figure}[h!] 
\begin{center}
\includegraphics[width=0.9\linewidth]{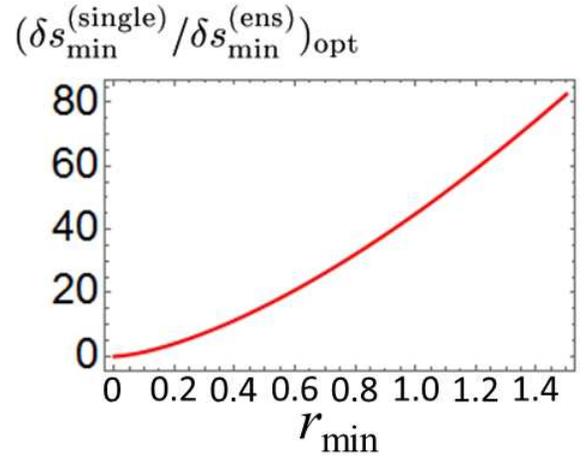} 
\caption{{{Plot of the optimized}} ratio $\delta s^{(\rm{single})}_{\rm{min}}/\delta
  s^{(\rm{ens})}_{\rm{min}}$ {{versus}} $r_{\rm{min}}$ where we {{chose}} $r_{\rm{max}}$ and
 $z_{\rm{max}}$ to maximize the ratio.
 Except {{for}} these two
 parameters, we used the same parameters as the Fig. \ref{cylthreed}.
 The ensemble {{of}} probe spins shows better performance than the single probe
 spin as long as $r_{\rm{min}}\geq 0.08\,\mu $m.
 }
  \label{cyltwod}
\end{center}
\end{figure}

We calculated the uncertainty of the estimation as
\begin{eqnarray}
 \delta s^{(\rm{ens})}_{\rm{min}}\simeq \frac{e^{\frac{1}{2}}\sqrt{2\rho \pi z_{\rm{max}}(r^2_{\rm{max}}-r^2_{\rm{min}})}}{|
  2G\rho  \int^{r_{\rm{max}}}_{r_{\rm{min}}} dr\int
  _0^{2\pi }d\theta  \int ^{z_{\rm{max}}}_{-z_{\rm{max}}}dz
 \frac{r(\frac{3z^2}{r^2+z^2}-1)}{(r^2+z^2)^{\frac{3}{2}}}
  \sqrt{\frac{T^{(\rm{ens})}_2T}{2}}|}.\nonumber 
\end{eqnarray}
For comparison, we consider {{the}} uncertainty when we use a single
probe spin for the spin detection by substituting
$x=r_{\rm{min}}$ and $y=z=0$  into 
Eq. \ref{single}.
\begin{eqnarray}
 \delta s^{(\rm{single})}_{\rm{min}}&\simeq
  &\frac{e^{\frac{1}{2}}(r_{\rm{min}})^{3} }{2G
  }\frac{1}{\sqrt{\frac{TT^{\rm{(single)}}_2}{2 } }}.\nonumber
\end{eqnarray}
We can define the ratio of the uncertainty of the estimation as
\begin{eqnarray}
 \frac{\delta s^{(\rm{single})}_{\rm{min}}}{\delta
  s^{(\rm{ens})}_{\rm{min}}}=
  \frac{ (r_{\rm{min}})^3\sqrt{\frac{T_2^{\rm{(ens)}}}{T_2^{\rm{(single)}}}}\int^{r_{\rm{max}}}_{r_{\rm{min}}} dr  \int ^{z_{\rm{max}}}_{-z_{\rm{max}}}dz
 \frac{r(\frac{3z^2}{r^2+z^2}-1)}{(r^2+z^2)^{\frac{3}{2}}}}{\sqrt{
 z_{\rm{max}}(r^2_{\rm{max}}-r^2_{\rm{min}})/(2\pi \rho )}}\ \ 
\end{eqnarray}

We plot the ratio $\delta s^{(\rm{single})}_{\rm{min}}/\delta
  s^{(\rm{ens})}_{\rm{min}}$ against $z_{\rm{max}}$ and $r_{\rm{max}}$
  in Fig. \ref{cylthreed}, where we fix
 $r_{\rm{min}}=1\,\mu $m.
  We have an optimal set of 
$z_{\rm{max}}$ and $r_{\rm{max}}$. The maximized ratio is {{approximately}}
  $\delta s^{(\rm{single})}_{\rm{min}}/\delta
  s^{(\rm{ens})}_{\rm{min}}\simeq 45${{;therefore,}} the sensitivity with the ensemble {{of}}
  probe spins is much better than that with the single probe spin. {{In addition}},
  we conclude that, if we use an ensemble of 
  probe spins, the cylinder
  configuration shows a better performance for the
  spin detection than the {{columnar}} configuration.
  Moreover, we {{plotted the}} an optimized value of $\delta s^{(\rm{single})}_{\rm{min}}/\delta
  s^{(\rm{ens})}_{\rm{min}}$ {{versus}} $r_{\rm{min}}$ in  Fig. \ref{cyltwod},
  where we chose
  $z_{\rm{max}}$ and $r_{\rm{max}}$ to maximize $\delta s^{(\rm{single})}_{\rm{min}}/\delta
  s^{(\rm{ens})}_{\rm{min}}$. The probe-spin ensemble in the {{cylindrical}} configuration
  {{has}} better sensitivity than the single probe spin as long as $r_{\rm{min}}\geq 0.08\ \mu $m.

      \begin{figure}[h!] 
\begin{center}
\includegraphics[width=0.9\linewidth]{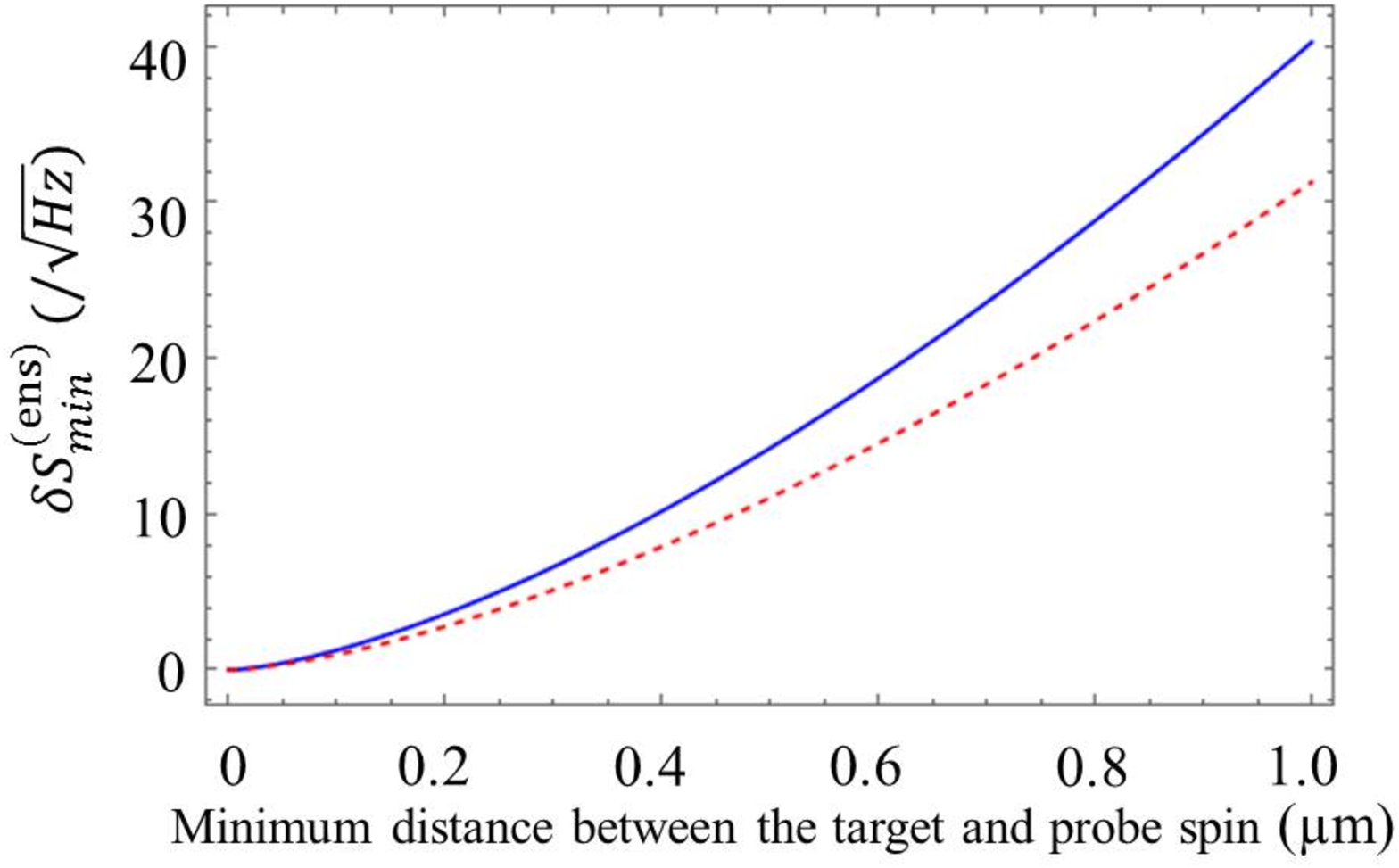} 
\caption{We plot an optimize ratio $\delta
  s^{(\rm{ens})}_{\rm{min}}$ against $z_{\rm{min}}$ ($r_{\rm{min}}$) to
 detect a single electron target spin
 for columnar (cylindrical)
  distribution of the probe spins by a
 continuous (dashed) line.
 Here, we choose $r_{\rm{max}}$ and
 $z_{\rm{max}}$ to maximize this ratio
 Except these two
 parameters, we used the same parameters as the Fig. \ref{cylthreed}.
 }
  \label{absolute}
\end{center}
\end{figure}

Finally, we plotted $\delta
  s^{(\rm{ens})}_{\rm{min}}$ to estimate the necessary time for the spin
  detection.
  We considered an electron spin to be the target spin. As shown in 
  Fig. \ref{absolute}, when the probe spins are distributed in a
   columnar (cylindrical) form, we obtain $\delta
  s^{(\rm{ens})}_{\rm{min}}=10$ for $z_{\rm{min}}\simeq 395$nm
  ($r_{\rm{min}}\simeq 468$nm) when we repeat the experiment for $T=1$ $s$. Since we need to achieve $\delta
  s^{(\rm{ens})}_{\rm{min}}\simeq 1$ to detect the target spin, the required
  time for the detection is approximately $T\simeq 100$ $s$. Therefore, when the distance between the target spin and
  probe spins is of the order of hundreds of nanometers, it takes around
  a few minutes
  to detect a single electron spin with the probe spin
  ensemble in our scheme, which is
  more than one order of magnitude
  faster
  than the case of a single probe spin as shown in Figs.
  \ref{coltwod} and \ref{cyltwod}.

\section{CONCLUSIONS}
{{We}} investigated {{the}} sensitivity {{of}} a single target
spin {{detection}} using an ensemble of probe spins. The use of {{a}} probe spin
ensemble increases the signal from the target spin while the
projection noise becomes more relevant as we increase the number of 
probe spins.
We {{demonstrated}} that, by optimizing the
distribution of the probe spins, the signal enhancement of the probe-
spin ensemble becomes much larger than the projection noise, which {{allows}} us
{{to}} detect the single target spin
 much more efficiently than {{in}} the case of a single probe spin.
{{In particular}}, our scheme is useful when the distance between {{the}} target spin
and {{the}} probe spins is {{of the order of}} hundreds of nanometers, which makes it
difficult for the single probe spin to detect the target due to the weak signal.
Our results pave the way for a new strategy
 to realize reliable single-spin detections. 

We thank Sayaka Kitazawa for useful discussions.
{{It}} was supported by JSPS KAKENHI Grant No.
15K17732. This work was also supported by MEXT
KAKENHI Grants No. 15H05868, No. 15H05870, No. 15H03996, No. 26220602 and No. 26249108.
The work was also supported by Advanced Photon Science Alliance (APSA), JSPS Core-to-Core Program, and Spin-NRJ.


\begin{thebibliography}{36}
\expandafter\ifx\csname natexlab\endcsname\relax\def\natexlab#1{#1}\fi
\expandafter\ifx\csname bibnamefont\endcsname\relax
  \def\bibnamefont#1{#1}\fi
\expandafter\ifx\csname bibfnamefont\endcsname\relax
  \def\bibfnamefont#1{#1}\fi
\expandafter\ifx\csname citenamefont\endcsname\relax
  \def\citenamefont#1{#1}\fi
\expandafter\ifx\csname url\endcsname\relax
  \def\url#1{\texttt{#1}}\fi
\expandafter\ifx\csname urlprefix\endcsname\relax\def\urlprefix{URL }\fi
\providecommand{\bibinfo}[2]{#2}
\providecommand{\eprint}[2][]{\url{#2}}

\bibitem[{\citenamefont{Ramsden}(2006)}]{ramsden2006hall}
\bibinfo{author}{\bibfnamefont{E.}~\bibnamefont{Ramsden}},
  \emph{\bibinfo{title}{Hall-effect sensors. newnes}} (\bibinfo{year}{2006}).

\bibitem[{\citenamefont{Vasyukov et~al.}(2013)\citenamefont{Vasyukov, Anahory,
  Embon, Halbertal, Cuppens, Neeman, Finkler, Segev, Myasoedov, Rappaport
  et~al.}}]{vasyukov2013scanning}
\bibinfo{author}{\bibfnamefont{D.}~\bibnamefont{Vasyukov}},
  \bibinfo{author}{\bibfnamefont{Y.}~\bibnamefont{Anahory}},
  \bibinfo{author}{\bibfnamefont{L.}~\bibnamefont{Embon}},
  \bibinfo{author}{\bibfnamefont{D.}~\bibnamefont{Halbertal}},
  \bibinfo{author}{\bibfnamefont{J.}~\bibnamefont{Cuppens}},
  \bibinfo{author}{\bibfnamefont{L.}~\bibnamefont{Neeman}},
  \bibinfo{author}{\bibfnamefont{A.}~\bibnamefont{Finkler}},
  \bibinfo{author}{\bibfnamefont{Y.}~\bibnamefont{Segev}},
  \bibinfo{author}{\bibfnamefont{Y.}~\bibnamefont{Myasoedov}},
  \bibinfo{author}{\bibfnamefont{M.~L.} \bibnamefont{Rappaport}},
  \bibnamefont{et~al.}, \bibinfo{journal}{Nature nanotechnology}
  \textbf{\bibinfo{volume}{8}}, \bibinfo{pages}{639} (\bibinfo{year}{2013}).

\bibitem[{\citenamefont{Bal et~al.}(2012)\citenamefont{Bal, Deng, Orgiazzi,
  Ong, and Lupascu}}]{bal2012ultrasensitive}
\bibinfo{author}{\bibfnamefont{M.}~\bibnamefont{Bal}},
  \bibinfo{author}{\bibfnamefont{C.}~\bibnamefont{Deng}},
  \bibinfo{author}{\bibfnamefont{J.}~\bibnamefont{Orgiazzi}},
  \bibinfo{author}{\bibfnamefont{F.}~\bibnamefont{Ong}}, \bibnamefont{and}
  \bibinfo{author}{\bibfnamefont{A.}~\bibnamefont{Lupascu}},
  \bibinfo{journal}{Nature Communications} \textbf{\bibinfo{volume}{3}},
  \bibinfo{pages}{1324} (\bibinfo{year}{2012}).

\bibitem[{\citenamefont{Toida et~al.}(2016)\citenamefont{Toida, Matsuzaki,
  Kakuyanagi, Zhu, Munro, Nemoto, Yamaguchi, and Saito}}]{toida2016electron}
\bibinfo{author}{\bibfnamefont{H.}~\bibnamefont{Toida}},
  \bibinfo{author}{\bibfnamefont{Y.}~\bibnamefont{Matsuzaki}},
  \bibinfo{author}{\bibfnamefont{K.}~\bibnamefont{Kakuyanagi}},
  \bibinfo{author}{\bibfnamefont{X.}~\bibnamefont{Zhu}},
  \bibinfo{author}{\bibfnamefont{W.~J.} \bibnamefont{Munro}},
  \bibinfo{author}{\bibfnamefont{K.}~\bibnamefont{Nemoto}},
  \bibinfo{author}{\bibfnamefont{H.}~\bibnamefont{Yamaguchi}},
  \bibnamefont{and} \bibinfo{author}{\bibfnamefont{S.}~\bibnamefont{Saito}},
  \bibinfo{journal}{Applied Physics Letters} \textbf{\bibinfo{volume}{108}},
  \bibinfo{pages}{052601} (\bibinfo{year}{2016}).

\bibitem[{\citenamefont{Bienfait et~al.}(2016)\citenamefont{Bienfait, Pla,
  Kubo, Stern, Zhou, Lo, Weis, Schenkel, Thewalt, Vion
  et~al.}}]{bienfait2016reaching}
\bibinfo{author}{\bibfnamefont{A.}~\bibnamefont{Bienfait}},
  \bibinfo{author}{\bibfnamefont{J.}~\bibnamefont{Pla}},
  \bibinfo{author}{\bibfnamefont{Y.}~\bibnamefont{Kubo}},
  \bibinfo{author}{\bibfnamefont{M.}~\bibnamefont{Stern}},
  \bibinfo{author}{\bibfnamefont{X.}~\bibnamefont{Zhou}},
  \bibinfo{author}{\bibfnamefont{C.}~\bibnamefont{Lo}},
  \bibinfo{author}{\bibfnamefont{C.}~\bibnamefont{Weis}},
  \bibinfo{author}{\bibfnamefont{T.}~\bibnamefont{Schenkel}},
  \bibinfo{author}{\bibfnamefont{M.}~\bibnamefont{Thewalt}},
  \bibinfo{author}{\bibfnamefont{D.}~\bibnamefont{Vion}}, \bibnamefont{et~al.},
  \bibinfo{journal}{Nature nanotechnology} \textbf{\bibinfo{volume}{11}},
  \bibinfo{pages}{253} (\bibinfo{year}{2016}).

\bibitem[{\citenamefont{Rugar et~al.}(2004)\citenamefont{Rugar, Budakian,
  Mamin, and Chui}}]{rugar2004single}
\bibinfo{author}{\bibfnamefont{D.}~\bibnamefont{Rugar}},
  \bibinfo{author}{\bibfnamefont{R.}~\bibnamefont{Budakian}},
  \bibinfo{author}{\bibfnamefont{H.}~\bibnamefont{Mamin}}, \bibnamefont{and}
  \bibinfo{author}{\bibfnamefont{B.}~\bibnamefont{Chui}},
  \bibinfo{journal}{Nature} \textbf{\bibinfo{volume}{430}},
  \bibinfo{pages}{329} (\bibinfo{year}{2004}).

\bibitem[{\citenamefont{Degen}(2008)}]{degen2008microscopy}
\bibinfo{author}{\bibfnamefont{C.}~\bibnamefont{Degen}},
  \bibinfo{journal}{Nature nanotechnology} \textbf{\bibinfo{volume}{3}},
  \bibinfo{pages}{643} (\bibinfo{year}{2008}).

\bibitem[{\citenamefont{Maze et~al.}(2008)\citenamefont{Maze, Stanwix, Hodges,
  Hong, Taylor, Cappellaro, Jiang, Dutt, Togan, Zibrov
  et~al.}}]{maze2008nanoscale}
\bibinfo{author}{\bibfnamefont{J.}~\bibnamefont{Maze}},
  \bibinfo{author}{\bibfnamefont{P.}~\bibnamefont{Stanwix}},
  \bibinfo{author}{\bibfnamefont{J.}~\bibnamefont{Hodges}},
  \bibinfo{author}{\bibfnamefont{S.}~\bibnamefont{Hong}},
  \bibinfo{author}{\bibfnamefont{J.}~\bibnamefont{Taylor}},
  \bibinfo{author}{\bibfnamefont{P.}~\bibnamefont{Cappellaro}},
  \bibinfo{author}{\bibfnamefont{L.}~\bibnamefont{Jiang}},
  \bibinfo{author}{\bibfnamefont{M.}~\bibnamefont{Dutt}},
  \bibinfo{author}{\bibfnamefont{E.}~\bibnamefont{Togan}},
  \bibinfo{author}{\bibfnamefont{A.}~\bibnamefont{Zibrov}},
  \bibnamefont{et~al.}, \bibinfo{journal}{Nature}
  \textbf{\bibinfo{volume}{455}}, \bibinfo{pages}{644} (\bibinfo{year}{2008}),
  ISSN \bibinfo{issn}{0028-0836}.

\bibitem[{\citenamefont{Taylor et~al.}(2008)\citenamefont{Taylor, Cappellaro,
  Childress, Jiang, Budker, Hemmer, Yacoby, Walsworth, and
  Lukin}}]{taylor2008high}
\bibinfo{author}{\bibfnamefont{J.}~\bibnamefont{Taylor}},
  \bibinfo{author}{\bibfnamefont{P.}~\bibnamefont{Cappellaro}},
  \bibinfo{author}{\bibfnamefont{L.}~\bibnamefont{Childress}},
  \bibinfo{author}{\bibfnamefont{L.}~\bibnamefont{Jiang}},
  \bibinfo{author}{\bibfnamefont{D.}~\bibnamefont{Budker}},
  \bibinfo{author}{\bibfnamefont{P.}~\bibnamefont{Hemmer}},
  \bibinfo{author}{\bibfnamefont{A.}~\bibnamefont{Yacoby}},
  \bibinfo{author}{\bibfnamefont{R.}~\bibnamefont{Walsworth}},
  \bibnamefont{and} \bibinfo{author}{\bibfnamefont{M.}~\bibnamefont{Lukin}},
  \bibinfo{journal}{Nature Physics} \textbf{\bibinfo{volume}{4}},
  \bibinfo{pages}{810} (\bibinfo{year}{2008}).

\bibitem[{\citenamefont{Balasubramanian
  et~al.}(2008)\citenamefont{Balasubramanian, Chan, Kolesov, Al-Hmoud, Tisler,
  Shin, Kim, Wojcik, Hemmer, Krueger et~al.}}]{balasubramanian2008nanoscale}
\bibinfo{author}{\bibfnamefont{G.}~\bibnamefont{Balasubramanian}},
  \bibinfo{author}{\bibfnamefont{I.}~\bibnamefont{Chan}},
  \bibinfo{author}{\bibfnamefont{R.}~\bibnamefont{Kolesov}},
  \bibinfo{author}{\bibfnamefont{M.}~\bibnamefont{Al-Hmoud}},
  \bibinfo{author}{\bibfnamefont{J.}~\bibnamefont{Tisler}},
  \bibinfo{author}{\bibfnamefont{C.}~\bibnamefont{Shin}},
  \bibinfo{author}{\bibfnamefont{C.}~\bibnamefont{Kim}},
  \bibinfo{author}{\bibfnamefont{A.}~\bibnamefont{Wojcik}},
  \bibinfo{author}{\bibfnamefont{P.}~\bibnamefont{Hemmer}},
  \bibinfo{author}{\bibfnamefont{A.}~\bibnamefont{Krueger}},
  \bibnamefont{et~al.}, \bibinfo{journal}{Nature}
  \textbf{\bibinfo{volume}{455}}, \bibinfo{pages}{648} (\bibinfo{year}{2008}).

\bibitem[{\citenamefont{Schaffry et~al.}(2011)\citenamefont{Schaffry, Gauger,
  Morton, and Benjamin}}]{schaffry2011proposed}
\bibinfo{author}{\bibfnamefont{M.}~\bibnamefont{Schaffry}},
  \bibinfo{author}{\bibfnamefont{E.}~\bibnamefont{Gauger}},
  \bibinfo{author}{\bibfnamefont{J.}~\bibnamefont{Morton}}, \bibnamefont{and}
  \bibinfo{author}{\bibfnamefont{S.}~\bibnamefont{Benjamin}},
  \bibinfo{journal}{Phys. Rev. Lett.} \textbf{\bibinfo{volume}{107}},
  \bibinfo{pages}{207210} (\bibinfo{year}{2011}).

\bibitem[{\citenamefont{M{\"u}ller et~al.}(2014)\citenamefont{M{\"u}ller, Kong,
  Cai, Melentijevi{\'c}, Stacey, Markham, Twitchen, Isoya, Pezzagna, Meijer
  et~al.}}]{muller2014nuclear}
\bibinfo{author}{\bibfnamefont{C.}~\bibnamefont{M{\"u}ller}},
  \bibinfo{author}{\bibfnamefont{X.}~\bibnamefont{Kong}},
  \bibinfo{author}{\bibfnamefont{J.-M.} \bibnamefont{Cai}},
  \bibinfo{author}{\bibfnamefont{K.}~\bibnamefont{Melentijevi{\'c}}},
  \bibinfo{author}{\bibfnamefont{A.}~\bibnamefont{Stacey}},
  \bibinfo{author}{\bibfnamefont{M.}~\bibnamefont{Markham}},
  \bibinfo{author}{\bibfnamefont{D.}~\bibnamefont{Twitchen}},
  \bibinfo{author}{\bibfnamefont{J.}~\bibnamefont{Isoya}},
  \bibinfo{author}{\bibfnamefont{S.}~\bibnamefont{Pezzagna}},
  \bibinfo{author}{\bibfnamefont{J.}~\bibnamefont{Meijer}},
  \bibnamefont{et~al.}, \bibinfo{journal}{Nature communications}
  \textbf{\bibinfo{volume}{5}} (\bibinfo{year}{2014}).

\bibitem[{\citenamefont{Staudacher et~al.}(2013)\citenamefont{Staudacher, Shi,
  Pezzagna, Meijer, Du, Meriles, Reinhard, and
  Wrachtrup}}]{staudacher2013nuclear}
\bibinfo{author}{\bibfnamefont{T.}~\bibnamefont{Staudacher}},
  \bibinfo{author}{\bibfnamefont{F.}~\bibnamefont{Shi}},
  \bibinfo{author}{\bibfnamefont{S.}~\bibnamefont{Pezzagna}},
  \bibinfo{author}{\bibfnamefont{J.}~\bibnamefont{Meijer}},
  \bibinfo{author}{\bibfnamefont{J.}~\bibnamefont{Du}},
  \bibinfo{author}{\bibfnamefont{C.~A.} \bibnamefont{Meriles}},
  \bibinfo{author}{\bibfnamefont{F.}~\bibnamefont{Reinhard}}, \bibnamefont{and}
  \bibinfo{author}{\bibfnamefont{J.}~\bibnamefont{Wrachtrup}},
  \bibinfo{journal}{Science} \textbf{\bibinfo{volume}{339}},
  \bibinfo{pages}{561} (\bibinfo{year}{2013}).

\bibitem[{\citenamefont{Mamin et~al.}(2013)\citenamefont{Mamin, Kim, Sherwood,
  Rettner, Ohno, Awschalom, and Rugar}}]{mamin2013nanoscale}
\bibinfo{author}{\bibfnamefont{H.}~\bibnamefont{Mamin}},
  \bibinfo{author}{\bibfnamefont{M.}~\bibnamefont{Kim}},
  \bibinfo{author}{\bibfnamefont{M.}~\bibnamefont{Sherwood}},
  \bibinfo{author}{\bibfnamefont{C.}~\bibnamefont{Rettner}},
  \bibinfo{author}{\bibfnamefont{K.}~\bibnamefont{Ohno}},
  \bibinfo{author}{\bibfnamefont{D.}~\bibnamefont{Awschalom}},
  \bibnamefont{and} \bibinfo{author}{\bibfnamefont{D.}~\bibnamefont{Rugar}},
  \bibinfo{journal}{Science} \textbf{\bibinfo{volume}{339}},
  \bibinfo{pages}{557} (\bibinfo{year}{2013}).

\bibitem[{\citenamefont{Ohashi et~al.}(2013)\citenamefont{Ohashi, Rosskopf,
  Watanabe, Loretz, Tao, Hauert, Tomizawa, Ishikawa, Ishi-Hayase, Shikata
  et~al.}}]{ohashi2013negatively}
\bibinfo{author}{\bibfnamefont{K.}~\bibnamefont{Ohashi}},
  \bibinfo{author}{\bibfnamefont{T.}~\bibnamefont{Rosskopf}},
  \bibinfo{author}{\bibfnamefont{H.}~\bibnamefont{Watanabe}},
  \bibinfo{author}{\bibfnamefont{M.}~\bibnamefont{Loretz}},
  \bibinfo{author}{\bibfnamefont{Y.}~\bibnamefont{Tao}},
  \bibinfo{author}{\bibfnamefont{R.}~\bibnamefont{Hauert}},
  \bibinfo{author}{\bibfnamefont{S.}~\bibnamefont{Tomizawa}},
  \bibinfo{author}{\bibfnamefont{T.}~\bibnamefont{Ishikawa}},
  \bibinfo{author}{\bibfnamefont{J.}~\bibnamefont{Ishi-Hayase}},
  \bibinfo{author}{\bibfnamefont{S.}~\bibnamefont{Shikata}},
  \bibnamefont{et~al.}, \bibinfo{journal}{Nano letters}
  \textbf{\bibinfo{volume}{13}}, \bibinfo{pages}{4733} (\bibinfo{year}{2013}).

\bibitem[{\citenamefont{Rugar et~al.}(2015)\citenamefont{Rugar, Mamin,
  Sherwood, Kim, Rettner, Ohno, and Awschalom}}]{rugar2015proton}
\bibinfo{author}{\bibfnamefont{D.}~\bibnamefont{Rugar}},
  \bibinfo{author}{\bibfnamefont{H.}~\bibnamefont{Mamin}},
  \bibinfo{author}{\bibfnamefont{M.}~\bibnamefont{Sherwood}},
  \bibinfo{author}{\bibfnamefont{M.}~\bibnamefont{Kim}},
  \bibinfo{author}{\bibfnamefont{C.}~\bibnamefont{Rettner}},
  \bibinfo{author}{\bibfnamefont{K.}~\bibnamefont{Ohno}}, \bibnamefont{and}
  \bibinfo{author}{\bibfnamefont{D.}~\bibnamefont{Awschalom}},
  \bibinfo{journal}{Nature nanotechnology} \textbf{\bibinfo{volume}{10}},
  \bibinfo{pages}{120} (\bibinfo{year}{2015}).

\bibitem[{\citenamefont{Vengalattore et~al.}(2007)\citenamefont{Vengalattore,
  Higbie, Leslie, Guzman, Sadler, and Stamper-Kurn}}]{vengalattore2007high}
\bibinfo{author}{\bibfnamefont{M.}~\bibnamefont{Vengalattore}},
  \bibinfo{author}{\bibfnamefont{J.}~\bibnamefont{Higbie}},
  \bibinfo{author}{\bibfnamefont{S.}~\bibnamefont{Leslie}},
  \bibinfo{author}{\bibfnamefont{J.}~\bibnamefont{Guzman}},
  \bibinfo{author}{\bibfnamefont{L.}~\bibnamefont{Sadler}}, \bibnamefont{and}
  \bibinfo{author}{\bibfnamefont{D.}~\bibnamefont{Stamper-Kurn}},
  \bibinfo{journal}{Physical review letters} \textbf{\bibinfo{volume}{98}},
  \bibinfo{pages}{200801} (\bibinfo{year}{2007}).

\bibitem[{\citenamefont{Kominis et~al.}(2003)\citenamefont{Kominis, Kornack,
  Allred, and Romalis}}]{kominis2003subfemtotesla}
\bibinfo{author}{\bibfnamefont{I.}~\bibnamefont{Kominis}},
  \bibinfo{author}{\bibfnamefont{T.}~\bibnamefont{Kornack}},
  \bibinfo{author}{\bibfnamefont{J.}~\bibnamefont{Allred}}, \bibnamefont{and}
  \bibinfo{author}{\bibfnamefont{M.}~\bibnamefont{Romalis}},
  \bibinfo{journal}{Nature} \textbf{\bibinfo{volume}{422}},
  \bibinfo{pages}{596} (\bibinfo{year}{2003}).

\bibitem[{\citenamefont{Acosta et~al.}(2009)\citenamefont{Acosta, Bauch,
  Ledbetter, Santori, Fu, Barclay, Beausoleil, Linget, Roch, Treussart
  et~al.}}]{acosta2009diamonds}
\bibinfo{author}{\bibfnamefont{V.}~\bibnamefont{Acosta}},
  \bibinfo{author}{\bibfnamefont{E.}~\bibnamefont{Bauch}},
  \bibinfo{author}{\bibfnamefont{M.}~\bibnamefont{Ledbetter}},
  \bibinfo{author}{\bibfnamefont{C.}~\bibnamefont{Santori}},
  \bibinfo{author}{\bibfnamefont{K.-M.} \bibnamefont{Fu}},
  \bibinfo{author}{\bibfnamefont{P.}~\bibnamefont{Barclay}},
  \bibinfo{author}{\bibfnamefont{R.}~\bibnamefont{Beausoleil}},
  \bibinfo{author}{\bibfnamefont{H.}~\bibnamefont{Linget}},
  \bibinfo{author}{\bibfnamefont{J.}~\bibnamefont{Roch}},
  \bibinfo{author}{\bibfnamefont{F.}~\bibnamefont{Treussart}},
  \bibnamefont{et~al.}, \bibinfo{journal}{Physical Review B}
  \textbf{\bibinfo{volume}{80}}, \bibinfo{pages}{115202}
  (\bibinfo{year}{2009}).

\bibitem[{\citenamefont{Maertz et~al.}(2010)\citenamefont{Maertz, Wijnheijmer,
  Fuchs, Nowakowski, and Awschalom}}]{maertz2010vector}
\bibinfo{author}{\bibfnamefont{B.}~\bibnamefont{Maertz}},
  \bibinfo{author}{\bibfnamefont{A.}~\bibnamefont{Wijnheijmer}},
  \bibinfo{author}{\bibfnamefont{G.}~\bibnamefont{Fuchs}},
  \bibinfo{author}{\bibfnamefont{M.}~\bibnamefont{Nowakowski}},
  \bibnamefont{and}
  \bibinfo{author}{\bibfnamefont{D.}~\bibnamefont{Awschalom}},
  \bibinfo{journal}{Applied Physics Letters} \textbf{\bibinfo{volume}{96}},
  \bibinfo{pages}{092504} (\bibinfo{year}{2010}).

\bibitem[{\citenamefont{Le~Sage et~al.}(2013)\citenamefont{Le~Sage, Arai,
  Glenn, DeVience, Pham, Rahn-Lee, Lukin, Yacoby, Komeili, and
  Walsworth}}]{le2013optical}
\bibinfo{author}{\bibfnamefont{D.}~\bibnamefont{Le~Sage}},
  \bibinfo{author}{\bibfnamefont{K.}~\bibnamefont{Arai}},
  \bibinfo{author}{\bibfnamefont{D.}~\bibnamefont{Glenn}},
  \bibinfo{author}{\bibfnamefont{S.}~\bibnamefont{DeVience}},
  \bibinfo{author}{\bibfnamefont{L.}~\bibnamefont{Pham}},
  \bibinfo{author}{\bibfnamefont{L.}~\bibnamefont{Rahn-Lee}},
  \bibinfo{author}{\bibfnamefont{M.}~\bibnamefont{Lukin}},
  \bibinfo{author}{\bibfnamefont{A.}~\bibnamefont{Yacoby}},
  \bibinfo{author}{\bibfnamefont{A.}~\bibnamefont{Komeili}}, \bibnamefont{and}
  \bibinfo{author}{\bibfnamefont{R.}~\bibnamefont{Walsworth}},
  \bibinfo{journal}{Nature} \textbf{\bibinfo{volume}{496}},
  \bibinfo{pages}{486} (\bibinfo{year}{2013}).

\bibitem[{\citenamefont{Wolf et~al.}(2015)\citenamefont{Wolf, Neumann,
  Nakamura, Sumiya, Ohshima, Isoya, and Wrachtrup}}]{wolf2015subpicotesla}
\bibinfo{author}{\bibfnamefont{T.}~\bibnamefont{Wolf}},
  \bibinfo{author}{\bibfnamefont{P.}~\bibnamefont{Neumann}},
  \bibinfo{author}{\bibfnamefont{K.}~\bibnamefont{Nakamura}},
  \bibinfo{author}{\bibfnamefont{H.}~\bibnamefont{Sumiya}},
  \bibinfo{author}{\bibfnamefont{T.}~\bibnamefont{Ohshima}},
  \bibinfo{author}{\bibfnamefont{J.}~\bibnamefont{Isoya}}, \bibnamefont{and}
  \bibinfo{author}{\bibfnamefont{J.}~\bibnamefont{Wrachtrup}},
  \bibinfo{journal}{Physical Review X} \textbf{\bibinfo{volume}{5}},
  \bibinfo{pages}{041001} (\bibinfo{year}{2015}).

\bibitem[{\citenamefont{Dolde et~al.}(2011)\citenamefont{Dolde, Fedder,
  Doherty, N{\"o}bauer, Rempp, Balasubramanian, Wolf, Reinhard, Hollenberg,
  Jelezko et~al.}}]{dolde2011electric}
\bibinfo{author}{\bibfnamefont{F.}~\bibnamefont{Dolde}},
  \bibinfo{author}{\bibfnamefont{H.}~\bibnamefont{Fedder}},
  \bibinfo{author}{\bibfnamefont{M.}~\bibnamefont{Doherty}},
  \bibinfo{author}{\bibfnamefont{T.}~\bibnamefont{N{\"o}bauer}},
  \bibinfo{author}{\bibfnamefont{F.}~\bibnamefont{Rempp}},
  \bibinfo{author}{\bibfnamefont{G.}~\bibnamefont{Balasubramanian}},
  \bibinfo{author}{\bibfnamefont{T.}~\bibnamefont{Wolf}},
  \bibinfo{author}{\bibfnamefont{F.}~\bibnamefont{Reinhard}},
  \bibinfo{author}{\bibfnamefont{L.}~\bibnamefont{Hollenberg}},
  \bibinfo{author}{\bibfnamefont{F.}~\bibnamefont{Jelezko}},
  \bibnamefont{et~al.}, \bibinfo{journal}{Nature Physics}
  \textbf{\bibinfo{volume}{7}}, \bibinfo{pages}{459} (\bibinfo{year}{2011}).

\bibitem[{\citenamefont{De~Lange et~al.}(2010)\citenamefont{De~Lange, Wang,
  Riste, Dobrovitski, and Hanson}}]{de2010universal}
\bibinfo{author}{\bibfnamefont{G.}~\bibnamefont{De~Lange}},
  \bibinfo{author}{\bibfnamefont{Z.}~\bibnamefont{Wang}},
  \bibinfo{author}{\bibfnamefont{D.}~\bibnamefont{Riste}},
  \bibinfo{author}{\bibfnamefont{V.}~\bibnamefont{Dobrovitski}},
  \bibnamefont{and} \bibinfo{author}{\bibfnamefont{R.}~\bibnamefont{Hanson}},
  \bibinfo{journal}{Science} \textbf{\bibinfo{volume}{330}},
  \bibinfo{pages}{60} (\bibinfo{year}{2010}).

\bibitem[{\citenamefont{Maletinsky et~al.}(2012)\citenamefont{Maletinsky, Hong,
  Grinolds, Hausmann, Lukin, Walsworth, Loncar, and
  Yacoby}}]{maletinsky2012robust}
\bibinfo{author}{\bibfnamefont{P.}~\bibnamefont{Maletinsky}},
  \bibinfo{author}{\bibfnamefont{S.}~\bibnamefont{Hong}},
  \bibinfo{author}{\bibfnamefont{M.}~\bibnamefont{Grinolds}},
  \bibinfo{author}{\bibfnamefont{B.}~\bibnamefont{Hausmann}},
  \bibinfo{author}{\bibfnamefont{M.}~\bibnamefont{Lukin}},
  \bibinfo{author}{\bibfnamefont{R.}~\bibnamefont{Walsworth}},
  \bibinfo{author}{\bibfnamefont{M.}~\bibnamefont{Loncar}}, \bibnamefont{and}
  \bibinfo{author}{\bibfnamefont{A.}~\bibnamefont{Yacoby}},
  \bibinfo{journal}{Nature Nanotechnology} \textbf{\bibinfo{volume}{7}},
  \bibinfo{pages}{320} (\bibinfo{year}{2012}).

\bibitem[{\citenamefont{Davies}(1994)}]{Go01a}
\bibinfo{author}{\bibfnamefont{G.}~\bibnamefont{Davies}},
  \emph{\bibinfo{title}{Properties and Growth of Diamond}}
  (\bibinfo{publisher}{Inspec/Iee}, \bibinfo{year}{1994}).

\bibitem[{\citenamefont{Gruber et~al.}(1997)\citenamefont{Gruber,
  Dr{\"a}benstedt, Tietz, Fleury, Wrachtrup, and
  Von~Borczyskowski}}]{gruber1997scanning}
\bibinfo{author}{\bibfnamefont{A.}~\bibnamefont{Gruber}},
  \bibinfo{author}{\bibfnamefont{A.}~\bibnamefont{Dr{\"a}benstedt}},
  \bibinfo{author}{\bibfnamefont{C.}~\bibnamefont{Tietz}},
  \bibinfo{author}{\bibfnamefont{L.}~\bibnamefont{Fleury}},
  \bibinfo{author}{\bibfnamefont{J.}~\bibnamefont{Wrachtrup}},
  \bibnamefont{and}
  \bibinfo{author}{\bibfnamefont{C.}~\bibnamefont{Von~Borczyskowski}},
  \bibinfo{journal}{Science} \textbf{\bibinfo{volume}{276}},
  \bibinfo{pages}{2012} (\bibinfo{year}{1997}).

\bibitem[{\citenamefont{Jelezko et~al.}(2002)\citenamefont{Jelezko, Popa,
  Gruber, Tietz, Wrachtrup, Nizovtsev, and Kilin}}]{jelezko2002single}
\bibinfo{author}{\bibfnamefont{F.}~\bibnamefont{Jelezko}},
  \bibinfo{author}{\bibfnamefont{I.}~\bibnamefont{Popa}},
  \bibinfo{author}{\bibfnamefont{A.}~\bibnamefont{Gruber}},
  \bibinfo{author}{\bibfnamefont{C.}~\bibnamefont{Tietz}},
  \bibinfo{author}{\bibfnamefont{J.}~\bibnamefont{Wrachtrup}},
  \bibinfo{author}{\bibfnamefont{A.}~\bibnamefont{Nizovtsev}},
  \bibnamefont{and} \bibinfo{author}{\bibfnamefont{S.}~\bibnamefont{Kilin}},
  \bibinfo{journal}{Appl. Phys. Lett.} \textbf{\bibinfo{volume}{81}},
  \bibinfo{pages}{2160} (\bibinfo{year}{2002}).

\bibitem[{\citenamefont{Jelezko et~al.}(2004)\citenamefont{Jelezko, Gaebel,
  Popa, Gruber, and Wrachtrup}}]{JGPGW01a}
\bibinfo{author}{\bibfnamefont{F.}~\bibnamefont{Jelezko}},
  \bibinfo{author}{\bibfnamefont{T.}~\bibnamefont{Gaebel}},
  \bibinfo{author}{\bibfnamefont{I.}~\bibnamefont{Popa}},
  \bibinfo{author}{\bibfnamefont{A.}~\bibnamefont{Gruber}}, \bibnamefont{and}
  \bibinfo{author}{\bibfnamefont{J.}~\bibnamefont{Wrachtrup}},
  \bibinfo{journal}{Phys. Rev. Lett} \textbf{\bibinfo{volume}{92}},
  \bibinfo{pages}{076401} (\bibinfo{year}{2004}).

\bibitem[{\citenamefont{Balasubramanian
  et~al.}(2009)\citenamefont{Balasubramanian, Neumann, Twitchen, Markham,
  Kolesov, Mizuochi, Isoya, Achard, Beck, Tissler
  et~al.}}]{balasubramanian2009ultralong}
\bibinfo{author}{\bibfnamefont{G.}~\bibnamefont{Balasubramanian}},
  \bibinfo{author}{\bibfnamefont{P.}~\bibnamefont{Neumann}},
  \bibinfo{author}{\bibfnamefont{D.}~\bibnamefont{Twitchen}},
  \bibinfo{author}{\bibfnamefont{M.}~\bibnamefont{Markham}},
  \bibinfo{author}{\bibfnamefont{R.}~\bibnamefont{Kolesov}},
  \bibinfo{author}{\bibfnamefont{N.}~\bibnamefont{Mizuochi}},
  \bibinfo{author}{\bibfnamefont{J.}~\bibnamefont{Isoya}},
  \bibinfo{author}{\bibfnamefont{J.}~\bibnamefont{Achard}},
  \bibinfo{author}{\bibfnamefont{J.}~\bibnamefont{Beck}},
  \bibinfo{author}{\bibfnamefont{J.}~\bibnamefont{Tissler}},
  \bibnamefont{et~al.}, \bibinfo{journal}{Nature materials}
  \textbf{\bibinfo{volume}{8}}, \bibinfo{pages}{383} (\bibinfo{year}{2009}).

\bibitem[{\citenamefont{Mizuochi et~al.}(2009)\citenamefont{Mizuochi, Neumann,
  Rempp, Beck, Jacques, Siyushev, Nakamura, Twitchen, Watanabe, Yamasaki
  et~al.}}]{mizuochi2009coherence}
\bibinfo{author}{\bibfnamefont{N.}~\bibnamefont{Mizuochi}},
  \bibinfo{author}{\bibfnamefont{P.}~\bibnamefont{Neumann}},
  \bibinfo{author}{\bibfnamefont{F.}~\bibnamefont{Rempp}},
  \bibinfo{author}{\bibfnamefont{J.}~\bibnamefont{Beck}},
  \bibinfo{author}{\bibfnamefont{V.}~\bibnamefont{Jacques}},
  \bibinfo{author}{\bibfnamefont{P.}~\bibnamefont{Siyushev}},
  \bibinfo{author}{\bibfnamefont{K.}~\bibnamefont{Nakamura}},
  \bibinfo{author}{\bibfnamefont{D.}~\bibnamefont{Twitchen}},
  \bibinfo{author}{\bibfnamefont{H.}~\bibnamefont{Watanabe}},
  \bibinfo{author}{\bibfnamefont{S.}~\bibnamefont{Yamasaki}},
  \bibnamefont{et~al.}, \bibinfo{journal}{Physical review B}
  \textbf{\bibinfo{volume}{80}}, \bibinfo{pages}{041201}
  (\bibinfo{year}{2009}).

\bibitem[{\citenamefont{Grezes et~al.}(2015)\citenamefont{Grezes, Julsgaard,
  Kubo, Ma, Stern, Bienfait, Nakamura, Isoya, Onoda, Ohshima
  et~al.}}]{grezes2015storage}
\bibinfo{author}{\bibfnamefont{C.}~\bibnamefont{Grezes}},
  \bibinfo{author}{\bibfnamefont{B.}~\bibnamefont{Julsgaard}},
  \bibinfo{author}{\bibfnamefont{Y.}~\bibnamefont{Kubo}},
  \bibinfo{author}{\bibfnamefont{W.}~\bibnamefont{Ma}},
  \bibinfo{author}{\bibfnamefont{M.}~\bibnamefont{Stern}},
  \bibinfo{author}{\bibfnamefont{A.}~\bibnamefont{Bienfait}},
  \bibinfo{author}{\bibfnamefont{K.}~\bibnamefont{Nakamura}},
  \bibinfo{author}{\bibfnamefont{J.}~\bibnamefont{Isoya}},
  \bibinfo{author}{\bibfnamefont{S.}~\bibnamefont{Onoda}},
  \bibinfo{author}{\bibfnamefont{T.}~\bibnamefont{Ohshima}},
  \bibnamefont{et~al.}, \bibinfo{journal}{Physical Review A}
  \textbf{\bibinfo{volume}{92}}, \bibinfo{pages}{020301}
  (\bibinfo{year}{2015}).

\bibitem[{\citenamefont{Bushev et~al.}(2011)\citenamefont{Bushev, Feofanov,
  Rotzinger, Protopopov, Cole, Wilson, Fischer, Lukashenko, and
  Ustinov}}]{bushev2011ultralow}
\bibinfo{author}{\bibfnamefont{P.}~\bibnamefont{Bushev}},
  \bibinfo{author}{\bibfnamefont{A.}~\bibnamefont{Feofanov}},
  \bibinfo{author}{\bibfnamefont{H.}~\bibnamefont{Rotzinger}},
  \bibinfo{author}{\bibfnamefont{I.}~\bibnamefont{Protopopov}},
  \bibinfo{author}{\bibfnamefont{J.}~\bibnamefont{Cole}},
  \bibinfo{author}{\bibfnamefont{C.}~\bibnamefont{Wilson}},
  \bibinfo{author}{\bibfnamefont{G.}~\bibnamefont{Fischer}},
  \bibinfo{author}{\bibfnamefont{A.}~\bibnamefont{Lukashenko}},
  \bibnamefont{and} \bibinfo{author}{\bibfnamefont{A.}~\bibnamefont{Ustinov}},
  \bibinfo{journal}{Physical Review B} \textbf{\bibinfo{volume}{84}},
  \bibinfo{pages}{060501} (\bibinfo{year}{2011}).

\bibitem[{\citenamefont{Tyryshkin et~al.}(2012)\citenamefont{Tyryshkin, Tojo,
  Morton, Riemann, Abrosimov, Becker, Pohl, Schenkel, Thewalt, Itoh
  et~al.}}]{tyryshkin2012electron}
\bibinfo{author}{\bibfnamefont{A.~M.} \bibnamefont{Tyryshkin}},
  \bibinfo{author}{\bibfnamefont{S.}~\bibnamefont{Tojo}},
  \bibinfo{author}{\bibfnamefont{J.~J.} \bibnamefont{Morton}},
  \bibinfo{author}{\bibfnamefont{H.}~\bibnamefont{Riemann}},
  \bibinfo{author}{\bibfnamefont{N.~V.} \bibnamefont{Abrosimov}},
  \bibinfo{author}{\bibfnamefont{P.}~\bibnamefont{Becker}},
  \bibinfo{author}{\bibfnamefont{H.-J.} \bibnamefont{Pohl}},
  \bibinfo{author}{\bibfnamefont{T.}~\bibnamefont{Schenkel}},
  \bibinfo{author}{\bibfnamefont{M.~L.} \bibnamefont{Thewalt}},
  \bibinfo{author}{\bibfnamefont{K.~M.} \bibnamefont{Itoh}},
  \bibnamefont{et~al.}, \bibinfo{journal}{Nature materials}
  \textbf{\bibinfo{volume}{11}}, \bibinfo{pages}{143} (\bibinfo{year}{2012}).

\bibitem[{\citenamefont{Tanaka et~al.}(2015)\citenamefont{Tanaka, Knott,
  Matsuzaki, Dooley, Yamaguchi, Munro, and Saito}}]{tanaka2015proposed}
\bibinfo{author}{\bibfnamefont{T.}~\bibnamefont{Tanaka}},
  \bibinfo{author}{\bibfnamefont{P.}~\bibnamefont{Knott}},
  \bibinfo{author}{\bibfnamefont{Y.}~\bibnamefont{Matsuzaki}},
  \bibinfo{author}{\bibfnamefont{S.}~\bibnamefont{Dooley}},
  \bibinfo{author}{\bibfnamefont{H.}~\bibnamefont{Yamaguchi}},
  \bibinfo{author}{\bibfnamefont{W.~J.} \bibnamefont{Munro}}, \bibnamefont{and}
  \bibinfo{author}{\bibfnamefont{S.}~\bibnamefont{Saito}},
  \bibinfo{journal}{Phys. Rev. Lett.} \textbf{\bibinfo{volume}{115}},
  \bibinfo{pages}{170801} (\bibinfo{year}{2015}).

\bibitem[{\citenamefont{Hadden et~al.}(2010)\citenamefont{Hadden, Harrison,
  Stanley-Clarke, Marseglia, Ho, Patton, OÇBrien, and
  Rarity}}]{hadden2010strongly}
\bibinfo{author}{\bibfnamefont{J.}~\bibnamefont{Hadden}},
  \bibinfo{author}{\bibfnamefont{J.}~\bibnamefont{Harrison}},
  \bibinfo{author}{\bibfnamefont{A.}~\bibnamefont{Stanley-Clarke}},
  \bibinfo{author}{\bibfnamefont{L.}~\bibnamefont{Marseglia}},
  \bibinfo{author}{\bibfnamefont{Y.-L.} \bibnamefont{Ho}},
  \bibinfo{author}{\bibfnamefont{B.}~\bibnamefont{Patton}},
  \bibinfo{author}{\bibfnamefont{J.}~\bibnamefont{OÇBrien}}, \bibnamefont{and}
  \bibinfo{author}{\bibfnamefont{J.}~\bibnamefont{Rarity}},
  \bibinfo{journal}{Applied Physics Letters} \textbf{\bibinfo{volume}{97}},
  \bibinfo{pages}{241901} (\bibinfo{year}{2010}).

\end{thebibliography}

\end{document}